\documentclass[journal, letterpaper]{IEEEtran}

\usepackage[top=1in, bottom=0.75in, left=0.75in, right=0.75in]{geometry}

\usepackage{mathtools}          
\usepackage{amssymb}
\usepackage{amsthm}

\usepackage{graphicx}
\graphicspath{{Figs/}}
\usepackage{booktabs}
\usepackage{array}
\usepackage{multirow}
\usepackage{arydshln}
\usepackage[caption=false,font=footnotesize]{subfig} 

\usepackage{tikz}
\usetikzlibrary{fit,positioning,automata,backgrounds,calc}

\usepackage[T1]{fontenc}
\usepackage[utf8]{inputenc}  
\usepackage{xcolor}
\usepackage{siunitx}
\usepackage{balance}
\usepackage{algorithm2e}
\usepackage{enumitem}

\usepackage{hyperref}
\hypersetup{
  colorlinks=true,
  linkcolor=blue,
  urlcolor=cyan,
  citecolor=blue,
}

\newcommand{\vectorstyle}[1]{\mathbf{#1}}

\usepackage{soul}

\usepackage{stackengine}

\renewcommand{\baselinestretch}{0.96}

\usepackage{etoolbox}
\AtBeginEnvironment{equation}{\vspace{-1pt}}
\AtEndEnvironment{equation}{\vspace{-1pt}}

\begin{document}


\title{\huge How to Capture Human Preference: Commissioning of a Robotic Use-Case via Preferential Bayesian Optimisation}


\author{
    S. De Witte$^{*}$, J. Taets$^{*}$, A. Retzler, G. Crevecoeur, and T. Lefebvre
    \thanks{$^{*}$These authors contributed equally to this work.}
    \thanks{All authors are with the Department of Electromechanical, Systems and Metal Engineering, Ghent University, B-9052 Ghent, Belgium (e-mail: {\tt\small \{sander.dewitte, jeroen.taets, andras.retzler, guillaume.crevecoeur, tom.lefebvre\}@ugent.be}).}
    \thanks{All authors are members of corelab MIRO, Flanders Make, Belgium.}
    \vspace*{-12pt}
}
\maketitle

\begin{abstract}
The popularity of Bayesian Optimization (BO) to automate or support the commissioning of engineering systems is rising. Conventional BO, however, relies on the availability of a scalar objective function. The latter is often difficult to define and rarely captures the nuanced judgement of expert operators in industrial settings. Preferential Bayesian Optimization (PBO) addresses this limitation by relying solely on pairwise preference feedback of a human expert, so-called duels. In this paper, we study PBO's capacity to commission a particular setup where a manipulator needs to push a block towards a target position. We benchmark state-of-the-art algorithms in both simulations and in the real world. Our results confirm that PBO can commission the set-up to the satisfaction of an expert operator whilst relying solely on binary preference feedback. To evaluate to what extend the same result can be achieved using conventional BO we investigate the experts decision consistency against an expert-designed cost function. Our study reveals that the experts fail to define a cost function that is in full agreement with their own decision process as witnessed in the PBO experiments. We then show that the auxiliary cost function that is constructed as a by-product of the PBO algorithms outperforms the expert-designed cost function in terms of decision consistency. Furthermore we demonstrate that this cost function can be used with conventional BO algorithms in an effort to reproduce the optimal design. This proofs the preference based cost function captures the experts' preferences perhaps more effectively than the experts could articulate preference themselves. In conclusion, we discuss downsides and propose directions for future research.
\end{abstract}

\section{Introduction} \label{sec:introduction}

It takes about 10 years to become an expert in a specific field. This rule of thumb was first claimed and later popularized by Simon and Chase \cite{ericsson1993role}\cite{chase1973chess}. Simon and his colleagues further concluded that expertise was the result of learning roughly 50,000 chunks of information. This expertise accumulates in the expert's mind and is hard to quantify, if at all. This makes experts invaluable and rare.

There is growing industrial interest in cognitive decision-support tools alleviating the challenges faced by junior employees \cite{lobo2025healthcare,gerlich2025ai}. Designing such tools requires both the extraction and quantification of expert knowledge, which has consequently attracted increasing attention from researchers. One approach to capturing expertise is to query operators through natural language. Several ontologies exist to formalize and transfer the qualitative knowledge obtained from such interactions with junior operators \cite{hao2021construction}. However, some expert skills are difficult to verbalize, let alone quantify. In management sciences, this type of know-how is referred to as tacit knowledge \cite{nonaka2009knowledge}. Among these skills, the ability to assess whether a given decision was favourable to the final outcome stands out as particularly essential.

In this work, we focus on the specific expert task of industrial commissioning, commonly referred to as tuning. Tuning generally involves an operator iteratively conducting on-site experiments to identify suitable machine settings that yield the best possible performance. The latter can include optimizing for power, efficiency, durability, or other unspecified goals. Tuning is required because changes in the operating context can degrade a system’s performance. These changes may range from variability in raw materials (e.g., in the textile industry: yarns and yarn types), variability in the environment of the system (e.g., operating in cold or warm weather), variability in the task (e.g., different weaving pattern), variability caused by the wear or ageing of the machine, leading to new or evolving interactions between components or subsystems (e.g., compressor interacting with a ventilation system).

\begin{figure}[t]
    \centering
    \subfloat[Before optimizing the parameters]{%
        \includegraphics[width=0.49\linewidth]{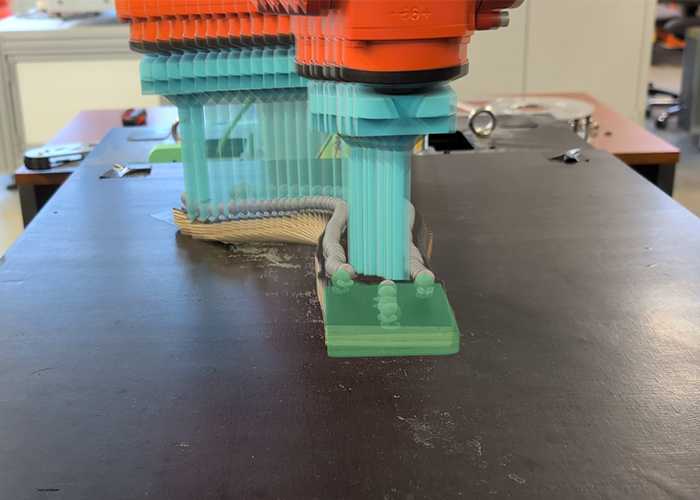}%
        \label{fig:exp_bad}}
    \hfil
    \subfloat[After optimizing the parameters]{%
        \includegraphics[width=0.49\linewidth]{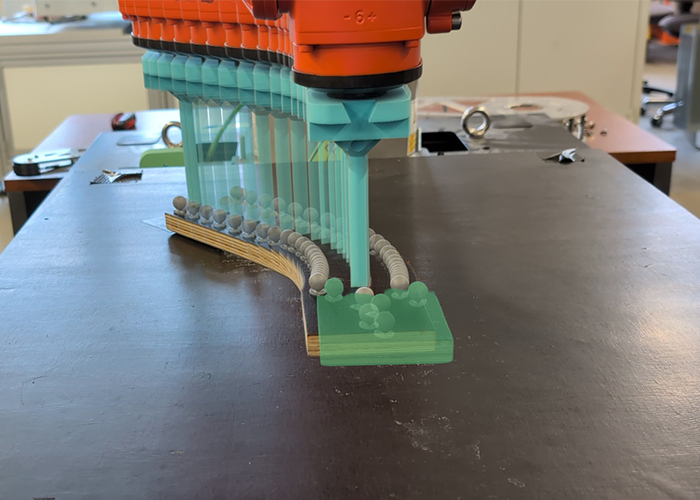}%
        \label{fig:exp_best_example}}
    \caption{Comparison of two experimental results on the physical setup, showing image sequences captured at 4~Hz during the robot motion. The control parameters for these runs are $\vectorstyle{\tau} = [4.0,\,1.0,\,0.31,\,0.1]$ and $\vectorstyle{\tau} = [1.0,\,3.45,\,0.14,\,2.0]$, and the motion takes approx. 20~s and 6~s, respectively. The latter corresponds to the best result obtained after Bayesian optimization. Top-view representations of these experiments are provided in Fig.~\ref{fig:example} and~\ref{fig:bo_benchmark}.}
    \label{fig:exp_best}
\end{figure}

The expertise of industrial operators often boils down to their ability to tune the system they operate to behave satisfactorily. During this process, the operator serves as both a discriminator and a sequential decision maker. The operator must accurately assess the extent to which further improvement is possible and determine the appropriate actions to achieve it. The main challenge is identifying the optimal settings with as few on-site experiments as possible, thereby minimizing downtime and material waste.

To automate or support this process, a cognitive tool must emulate key aspects of human decision-making, such as:
\begin{itemize}
    \item Humans have access to multimodal sensory information that extends beyond what can typically be captured and processed digitally. Even in highly instrumented facilities, not all sensory cues are represented, though they may be essential for judging whether the system operates satisfactorily.
    \item It is assumed that humans rely on mental models. These models may span both how settings translate into output features and how features translate into performance. Especially the latter is hard to verbalise let alone quantify.
    \item Experts follow a strategy for curating the next experiment. This strategy balances exploration and exploitation; some experiments are intended to improve context understanding, while others aim to discover optimal settings.
\end{itemize}

Managing all of these aspects in a single framework is challenging, though not impossible. One emerging paradigm is Bayesian Optimization (BO). BO has originally been conceived to solve black-box optimization problems with costly function calls, e.g. in the scope of Computer Aided Design (CAD) \cite{mockus2005bayesian,jones1998efficient}. Since then, BO has found widespread use in real-world settings, where optimization is performed directly through experiments rather than simulations.

Numerous recent studies have evaluated the capacity of BO to commission control systems \cite{shahriari2015taking,greenhill2020bayesian,coutinho2023bayesian,neumann2019data,konig2023safe,konig2025adaptive}. Vanilla BO algorithms, that are tailored to pure black-box optimization, only require that a numerical objective function is available. Members of this family may assist the decision process; however, they only replicate the human-based tuning conditions partially: (1) they require that the performance can be expressed numerically, meaning that an expert must define a cost function that reflects their preferences, and (2) they assume no prior knowledge about the system’s behaviour. Numerous extensions of standard BO algorithms have been proposed to overcome these limitations.

To incorporate prior knowledge, BO methods have been designed that make use of a simulator \cite{letham2019bayesian,eugene2023learning,sabbatella2024bayesian,karkaria2024towards,gafurov2025ai,nobar2024guided}. Particular to these approaches is that the simulator is updated based on on-site experiments.  The use of a simulator mimics the use of a mental process model, though this aspect is not pursued further in this work.

In this work, we take particular interest in the first problem. To address the absence of a cost function, various Preferential Bayesian Optimization (PBO) algorithms exist. In every iteration, an expert provides feedback to the algorithm by expressing a preference over two possible experiments, a duel \cite{EP_TS_MUC, chu2005preference}. This problem extends to applications beyond control tuning. The number of methods has exploded over the past years, followed by many benchmark studies. A limitation of these benchmarks is that they often simulate the outcome of a duel by evaluating an underlying cost function, thereby avoiding human input. Only a limited number of studies provide evidence that these methods function when a human operator is involved \cite{maccarini2022preference,tucker2022polar}.

The contributions of this work are threefold. 
\begin{itemize}
    \item We compare the performance of SOTA PBO algorithms both in simulation and practice on a robotic use-case.
    \item We investigate whether human preference can be replicated by a complicated engineered cost function devised by experts. We further demonstrate that a data-driven cost function can be extracted from the duels, which in turn can be used to commission the system using standard BO.
    \item We discuss downsides and possible future improvements of PBO based on our hands-on experience.
\end{itemize} 

\section{Preferential Bayesian Optimization}
The preference of a decision maker is modelled by a latent objective function $f:\mathcal{X} \mapsto \mathbb{R}$, $\mathcal{X}\subset\mathbb{R}^n$ denotes the continuous input domain. The goal is to identify $x^*$ corresponding to the most preferred conditions. 
\begin{equation}
	x^* = \arg\min_x f(x)
\end{equation}

In contrast to vanilla BO, the objective function $f$ cannot be queried directly to obtain scalar function values. Instead, information about $f$ is obtained indirectly through preferential feedback between pairs of inputs. Each query returns a binary relation of the form $x'' \succ x'$, signifying that option $x''$ is preferred over $x'$ by the oracle (e.g., a human user or an implicit reward process), implying $f(x'') < f(x')$. The goal in PBO is to identify $x^*$ using only these observations.


\subsection{Posterior modelling}

Many PBO algorithms construct a probabilistic estimate of the latent cost function, $f$ \cite{LA_EI}.  

To that end one assumes a Gaussian process (GP) prior with zero mean and stationary kernel $k : \mathcal{X} \times \mathcal{X} \to \mathbb{R}$
\begin{equation}
f \sim \mathcal{GP}(0, k(x, x'))
\end{equation}

As opposed to conventional BO, the training data $\mathcal{D}$ consist of $N$ pairwise comparisons, typically represented as ordered pairs $(x_{i,w}, x_{i,l})$ so that $x_{i,w} \succ x_{i,l}$. It is implied that the associated latent utilities satisfy, where $\varepsilon_{i,w}$ and $ \varepsilon_{i,l} \sim \mathcal{N}(0, \sigma_\text{noise}^2)$
\begin{equation}
f(x_{i,w}) + \varepsilon_{i,w} > f(x_{i,l}) + \varepsilon_{i,l},
\end{equation}

This condition can be expressed equivalently through the auxiliary variable $v_{i}$. By definition the it holds that $v_i < 0$ \begin{equation}
    v_i = f(x_{i,l}) + \varepsilon_{i,l} - f(x_{i,w}) - \varepsilon_{i,w} < 0
\end{equation}

Now assuming a data set exist of $N$ pairwise comparisons $\mathcal{D} = \{(x_{i,w},x_{i,l}),v_i\}_{i=1}^N$. The posterior output probability $f_\text{tes} = (f(x_{1,\text{tes}}), \dots, f(x_{m,\text{tes}}))^\top$, 
\begin{equation}
p(f_\text{tes} \mid v_t < 0)
= \frac{\Pr(v_t < 0 \mid f_\text{tes})\, p(f_\text{tes})}{\Pr(v_t < 0)}.
\end{equation}

Since the exact posterior is non-Gaussian, approximate inference is typically employed, such as Laplace's method, expectation propagation, or variational inference.

The resulting posterior provides a continuous estimate of the latent preference landscape over $\mathcal{X}$. This estimate can then be used to design acquisition functions that suggest informative new comparisons—balancing exploration of uncertain regions with exploitation of promising candidates—to efficiently converge toward the globally most preferred configuration.
	


\subsection{Acquisition functions} \label{sec:acquisition}
Given the posterior, an acquisition function is minimzed to determine the next duel which will be used to update the posterior. This procedure is repeated until convergence.

The Expected Improvement (\textit{EI}) and the Upper Confidence Bound (\textit{UCB}) are common acquisition functions for vanilla BO. These are maximized to find a single, most desirable point to sample next. PBO uses specialized acquisition strategies to select two different points in the parameter space that are informative to compare.

In \cite{LA_EI}, they choose to approximate the posterior with Laplace's method, and the incumbent is chosen as the first sample (i.e., the point with the best objective value on the approximated posterior), while the second sample is picked by \textit{EI}.
Similarly, in \cite{EP_MUC}, the incumbent is selected as a first point. However, the second point is chosen to maximize the epistemic uncertainty.  This is called the Maximally Uncertain Challenge (\textit{MUC}) acquisition strategy.

The limitations of Laplace's method were shown in \cite{DuelTSUCB}: it was discovered that the true posterior of the preference function is a skew GP, which is poorly approximated by Laplace's method due to its highly skewed pairwise marginals. 
The skew GP posterior can be calculated using a Markov Chain Monte Carlo (\textit{MCMC}) approach. Several duelling acquisition functions were proposed in \cite{DuelTSUCB}: the duelling Thompson, the duelling UCB, and the duelling Expected probability of improvement combined with Information Gain (\textit{EEIG}). All of these have two inputs, one for each of the points participating in a duel. They fix the first input to the sample corresponding to the best experiment seen so far, and globally optimize over the second input. Duelling acquisition functions with two inputs also appear in other works, such as the Expected Utility of the Best Option (\textit{EUBO}), introduced in \cite{EUBO}.

As an extension to \cite{DuelTSUCB}, the work \cite{HB} proposes the Hallucination Believer (\textit{HB}) method, where they use a modified posterior that is conditioned by a random sample from the original posterior itself. This sample is called hallucination. This method makes use of traditional acquisition functions, such as EI and UCB, making it possible to calculate them computationally efficiently for the PBO case.

Most of these works do not report on actual humans comparing experiments, instead they evaluate the methods on problems with a known cost function, i.e., they build the pairwise comparison data for PBO based on evaluations of the known cost function.

\section{Robotic use-case}
We consider a planar pushing problem to assess the use of Preferential Bayesian Optimization (PBO) on a relevant control task, where a finger-like robotic end-effector pushes an object toward a goal position. This task forms a well-defined yet challenging benchmark for human-in-the-loop tuning: it involves non-linear contact dynamics, frictional uncertainty, and a trade-off between speed and stability. The controller’s behaviour depends on a small set of tunable parameters, making it suitable for iterative optimization. Each trial is short, repeatable, and visually interpretable, allowing a human operator to express intuitive preferences between trajectories, such as smoother versus faster pushes. These properties make the pushing task an ideal testbed for evaluating how PBO incorporates subjective human judgments into controller tuning.

\subsection{Description of the setup}
\label{sec:Description}

\begin{figure}
\centering
\includegraphics[width=0.55\linewidth]{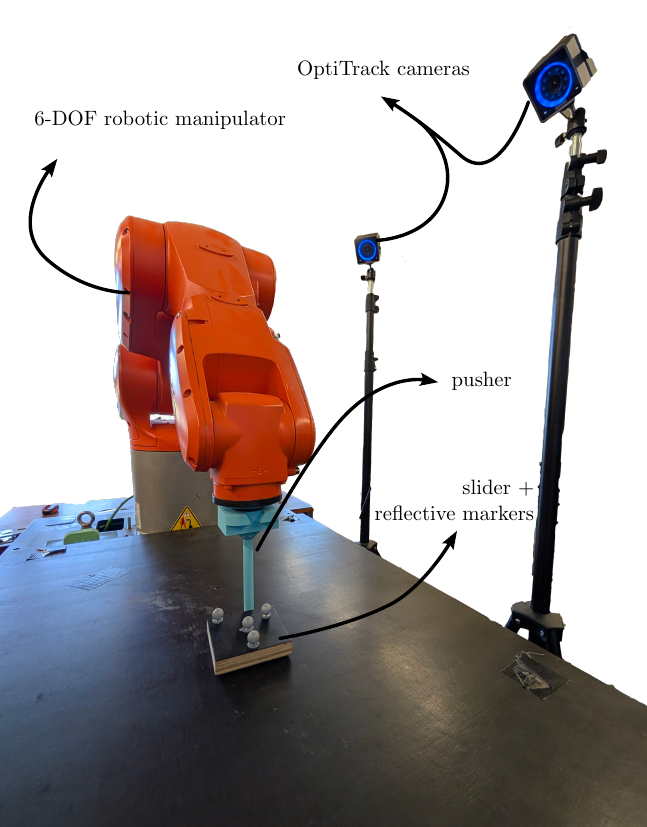}
\caption{Experimental setup with a finger-like end-effector attached to a KUKA KR AGILUS robot. Four OptiTrack cameras track a square block.}
\vspace{-12pt}
\label{fig:setup}
\end{figure}

The setup comprises a KUKA KR AGILUS (KR 6 R700-2) manipulator equipped with a finger-like end-effector to push a square block. An OptiTrack motion capture system provides perception with four Prime\textsuperscript{x} 13 cameras operated via the Motive software. The block is instrumented with four reflective markers, allowing accurate pose estimation (±0.20 mm) at 120 fps.

Each experiment involves pushing the block from a fixed initial state to a goal position $[x_d, y_d]=[0,30]$ cm. The pusher, the slider, and their interaction form the modelled system, with the square block as the slider and the cylindrical end-effector as the pusher. Motion is constrained to the plane, assuming continuous contact throughout.

The kinematic state of the slider and the control input of the pusher are represented by the vectors $\vectorstyle{x}$ and $\vectorstyle{u}$, respectively:
\begin{equation}
\vectorstyle{x} =
\begin{pmatrix}
x & y & \theta & d
\end{pmatrix}^\top, \qquad
\vectorstyle{u} =
\begin{pmatrix}
u_x & u_y
\end{pmatrix}^\top.
\end{equation}
Here, $x$ and $y$ denote the Cartesian coordinates of the slider’s centre in the global reference frame, $\theta$ represents its planar orientation, and $d$ specifies the contact point of the pusher along the lateral edge. The control input $\vectorstyle{u}$ corresponds to the pusher velocity expressed in the local frame, where $u_x$ and $u_y$ are the normal and tangential velocity components at the slider edge, respectively.

\begin{figure}[!t]
    \centering
    \includegraphics[width=0.5\columnwidth]{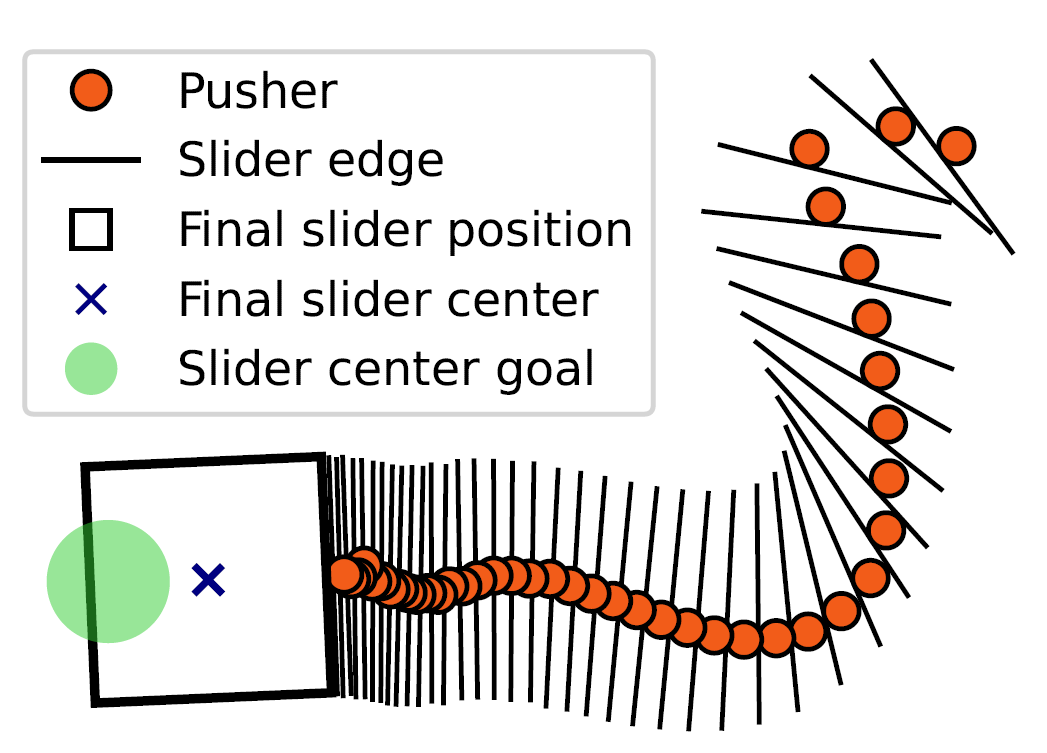}
    \vspace{-10pt}
    \caption{An example of a pusher–slider experiment. The pusher is continuously in contact with the slider, and pushes it forward until the centre of the slider either falls on the green region, or the push fails. The latter happen if the slider goes out of bounds, or the time limit has been reached. Such an out-of-time push is shown on this figure.}
    \vspace{-10pt}
    \label{fig:example}
\end{figure}

A schematic rendering of an experiment, with inadequate control parameters, is shown in Fig.~\ref{fig:example}. After a push, the block is manually placed in roughly the same spot. This results in some variability in the initial state, quantified by $\vectorstyle{x}_{\text{init}} = (4.74 \pm 0.32, -11.43 \pm 0.31, -50.19 \pm 1.41, 0 \pm 0)$ (mean ± standard deviation over all 630 experiments), where the position is given in centimetres and the angle in degrees.

\subsection{Control strategy}
We employ the cascaded quasi-static feedback controller originally proposed in \cite{dewitte2025}, which explains the derivation of the model and controller equations. The controller is based on a quasi-static model of planar pushing, which assumes that inertial effects are negligible compared to frictional forces and that contact friction at the pusher–slider interface is small relative to friction with the ground.

The controller consists of three nested feedback loops that regulate (i) the slider position, (ii) its orientation, and (iii) the pusher–slider contact offset. Each loop enforces a critically damped second-order error decay law, with dynamics determined by a characteristic time constant $\tau_i$.

The feedback gains are defined as
\begin{equation}
K_{p,i} = \frac{1}{\tau_i^2}, \qquad K_{d,i} = \frac{2}{\tau_i},
\end{equation}
resulting in four tunable parameters, $\vectorstyle{\tau} = (\tau_x, \tau_y, \tau_\theta, \tau_d)^\top$. These parameters determine the time scales of the cascaded loops, where the inner loops are typically tuned to be faster than the outer ones to ensure smooth and stable motion.

\subsection{Success criteria and performance metrics}
A push is considered successful if the centre of the block reaches the goal position within a radius of 2~cm, or if the maximum number of iterations $N_{\text{max}}$ is reached. The controller operates at a fixed sampling period of $\Delta t = 0.1$~s. 

From empirical observation, several qualitative features of a ``good'' push are characterized by the expert:
\begin{itemize}
    \item the slider reaches its goal position,
    \item the push is executed as fast as possible while maintaining continuous contact,
    \item and the pusher adapts smoothly to changes in the slider orientation, avoiding abrupt lateral motions.
\end{itemize}

The first two criteria can be quantified by the final distance to the goal and the total execution time, respectively. The last feature is captured through the lateral velocity of the pusher,
\begin{equation}
    v_{d,k} = \frac{d_k - d_{k-1}}{\Delta t},
\end{equation}
where $d_k$ denotes the signed lateral displacement of the contact point along the slider edge at time step $k$. These three aspects are reflected in the following cost terms:
\begin{align}
    J_1 &= (N - 1)\,\Delta t, \\[4pt]
    J_2 &= 
    \left\lVert
        \begin{pmatrix}
            x_N \\[2pt] y_N
        \end{pmatrix}
        -
        \begin{pmatrix}
            x_d \\[2pt] y_d
        \end{pmatrix}
    \right\rVert^2, \\[6pt]
    J_3 &= 
    \sum_{k=1}^{N-1} 
    \left(\frac{d_k - d_{k-1}}{\Delta t}\right)^2 \Delta t,
\end{align}
where $N$ denotes the total number of time steps. The first term $J_1$ represents the total push duration, $J_2$ penalizes the terminal position error, and $J_3$ penalizes rapid changes in the pusher’s lateral motion, promoting smooth trajectories.

\section{Experiments}

\subsection{Comparison of PBO approaches in simulations}

\begin{figure}[!t]
    \centering
    \includegraphics[width=\columnwidth]{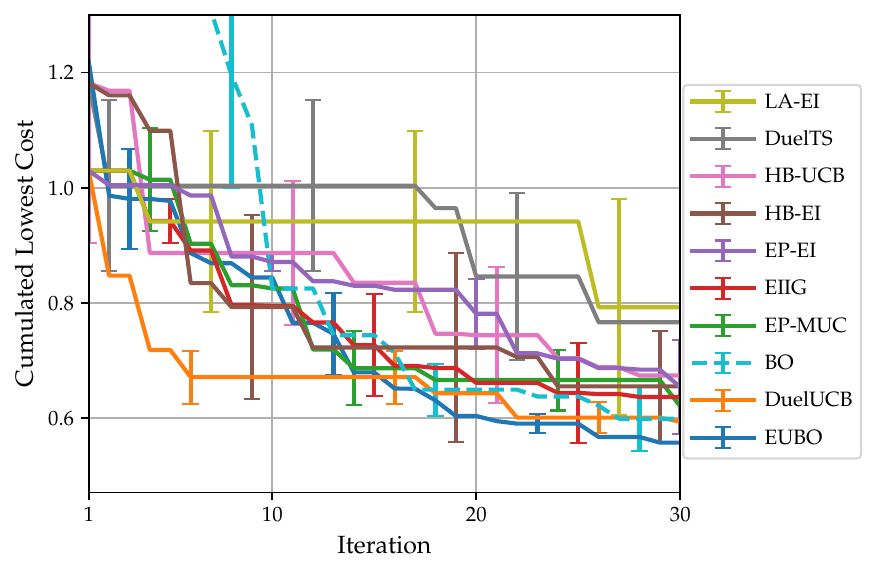}
    \vspace{-18pt}
    \caption{Average and standard error of the cumulated lowest cost in a simulation of the pusher–slider using state-of-the-art PBO methods. The actual cost shown on the graph was not given to the PBO methods: they solely used data on pairwise preferences.}
    \label{fig:simulation_benchmark}
\end{figure}
\begin{figure}[!t]
    \centering
    \includegraphics[width=\columnwidth]{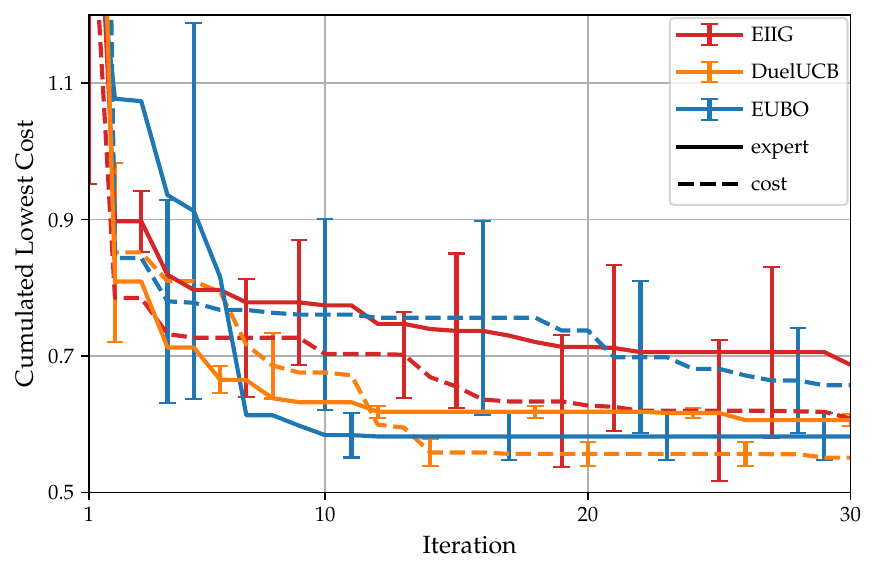}
    \vspace{-18pt}
    \caption{Average and standard error of the cumulative minimum cost obtained from experiments on the robotic setup. The y-axis represents the expert-defined cost function. The solid line corresponds to selections made by the human expert, while the dotted line indicates selections based on the lowest expert cost value.}
    \vspace{-8pt}
    \label{fig:experimental_benchmark}
\end{figure}

Although PBO inherently operates without access to an explicit cost function, in simulation, we introduce one to emulate a consistent expert. By defining a cost function that quantitatively reflects expert judgment, we ensure unbiased comparisons between algorithms. Following the same principle as in related work, this cost function is solely used to decide between two trials during simulation studies. The quality of an experiment is a weighted sum of three metrics
\begin{equation} \label{eq:cost_combination}
J = a_1 J_1 + a_2 J_2 + a_3 J_3.
\end{equation}

The weights $\vectorstyle{a} = (a_1, a_2, a_3)^\top = (0.1,1,1)^\top$ were selected based on some preliminary simulations by the expert to balance execution time, terminal accuracy, and smoothness of motion. This cost function thus represents the latent cost that PBO aims to minimize through preference feedback.

Each simulation experiment consists of a complete push starting from a fixed initial pose. The system dynamics follow the quasi-static model given in \cite{dewitte2025}, with Gaussian noise added on both the block pose and the input, to emulate real-world uncertainty. The PBO algorithms are given access only to pairwise preferences between trajectories based on their respective cost value, without direct access to the underlying cost $J$. The control parameters are explored within the range of $[1.0,\,4.0]$~seconds for the first two control parameters, and of $[0.1,\,2.0]$~seconds for the last two.

Each method was run for 15 preference-based iterations, starting from an identical prior consisting of 12 random initial samples. The entire procedure was repeated three times with different random seeds, though the same seeds were used for the different methods. Performance is reported as the mean and the standard deviation of the cumulative minimum cost over iterations.

The results are shown in Fig.~\ref{fig:simulation_benchmark}. The benchmark was implemented using the publicly available code from \cite{HB} and \cite{EUBO}, and the tested algorithms correspond to those that were earlier described in Section~\ref{sec:acquisition}. Overall, the performance is comparable across most methods; however, \textit{EUBO} and \textit{DuelUCB} exhibit the highest consistency, as reflected by their lower standard errors. For reference, we also include a comparison with vanilla BO, which shows lower-than-expected performance, likely due to excessive exploration in this setting.

\subsection{Comparison of PBO approaches on the set-up} \label{sec:PBO_setup}

Based on the simulation experiments, \textit{EUBO}, \textit{EIIG}, and \textit{DuelUCB} were selected for a comparative benchmark on the real setup. In contrast to the simulation experiments, the preferred push in each pairwise comparison was chosen by an expert rather than by a predefined cost function. Each benchmark run consisted of two initial random comparisons followed by 13 algorithm-driven ones, resulting in a total of 30 experiments per trial. While this number is considerably lower than that in the simulation, it still represents a realistic amount of interaction, roughly equivalent to the number of tuning steps an expert would typically perform.

To emulate the decision process of the expert in simulation, the initial cost function weights were designed to approximate human preferences. However, based on the experimental results, it was observed that these weights did not fully capture the actual priorities of the expert. In practice, the expert favoured low aggressiveness and accurate goal reaching over minimizing execution time, aiming to avoid locally aggressive minima. To better reflect these preferences, the cost function in Eq.~\ref{eq:cost_combination} was retrofitted with updated weights $\vectorstyle{a} = (0.035,7.5,11.0)$, rescaled to match the magnitude of the previous objective values. The refinement was performed by tuning the weights such that the experiments rated highest by the expert also corresponded to the lowest cost values under the new formulation. In addition to the expert-guided trials, experiments using this refined cost function as the selection criterion were conducted under identical settings and algorithms.

Figure~\ref{fig:experimental_benchmark} summarizes the experimental results. Solid lines represent the performance of the expert-guided experiments, while dotted lines indicate outcomes evaluated using the refined cost function. The expert-guided experiments appear slightly worse than those guided by the refined cost function; however, this discrepancy arises because the y-values of the dotted curves correspond to the expert-designed refined cost, which still does not fully capture the expert’s true preferences. Despite this, the overall trends remain consistent with the simulation results: \textit{EUBO} achieves the best performance, followed by \textit{DuelUCB} and \textit{EIIG}.

\begin{figure}[!t]
    \centering
    \includegraphics[width=0.8\columnwidth]{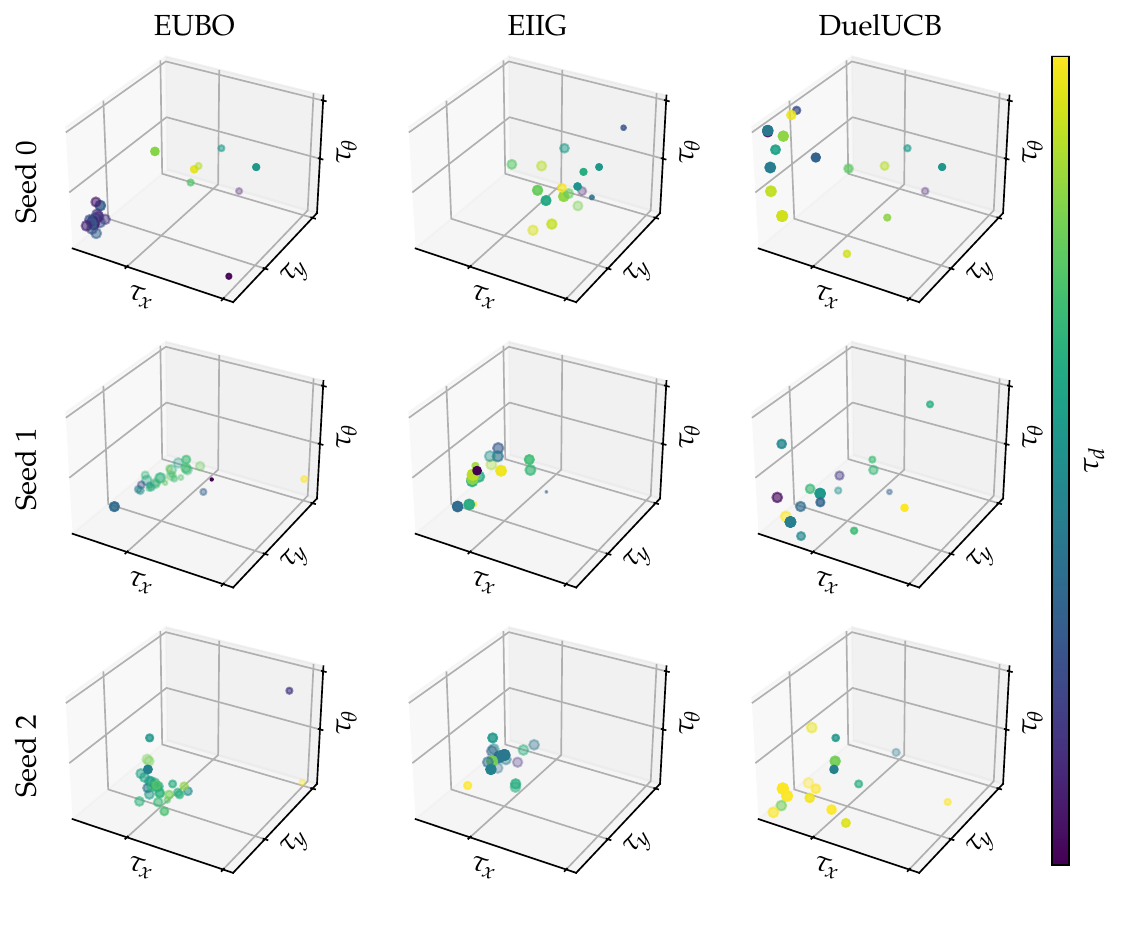}
    \vspace{-12pt}
    \caption{Exploration of the algorithms in the parameter space for the three seeds of the experiments on the set-up. The value of the fourth parameter is indicated by the colour.}
    \vspace{-8pt}
    \label{fig:experimental_exploration}
\end{figure}

Based on these experiments, several practical observations can be made. First, \textit{DuelUCB} compares the current best experiment with a newly sampled one. Although this strategy performs well in this setting, the newly sampled point was often located close to the current best, which limited overall exploration. Second, the tested algorithms occasionally became stuck at the boundary of the optimization space for one of the four control parameters. When this occurred, the algorithm consistently proposed boundary values in subsequent iterations (e.g., $\tau_x$ in seeds 0 and 1 for \textit{EUBO}, as shown in Fig.~\ref{fig:experimental_exploration}). Finally, the figure illustrates that the balance between exploration and exploitation differs substantially across algorithms, and adjusting this trade-off in practice remains challenging.

\subsection{Comparison of expert cost function and expert human selection}

\begin{figure*}[!t]
    \centering
    \vspace{-8pt}
    \includegraphics[width=0.6\textwidth]{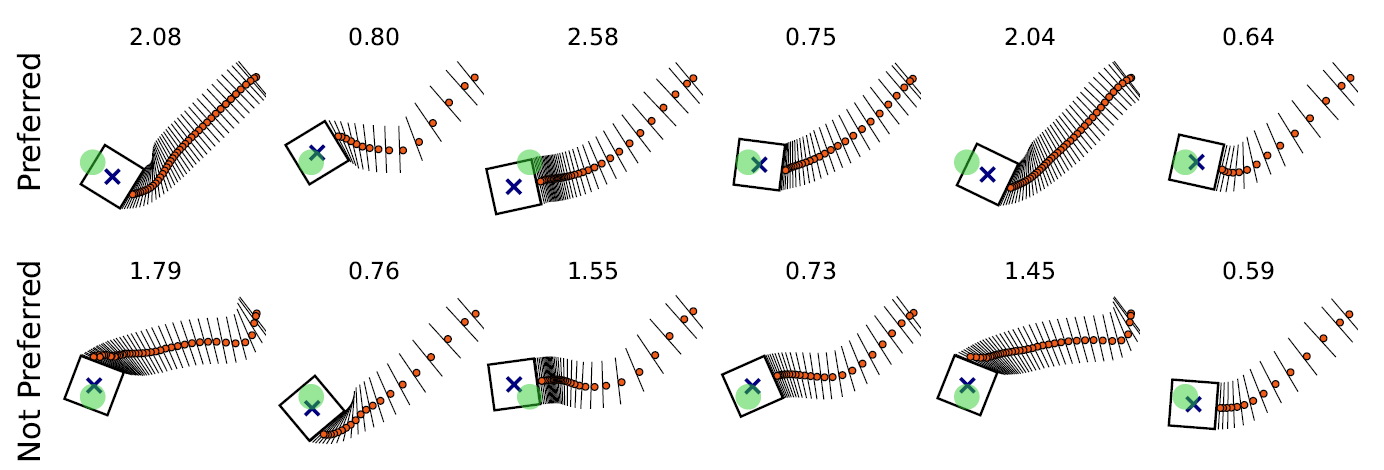}
    \caption{Selected experiments where human preference differed from the refined cost function. Each column shows the human-preferred experiment on top and the other (with a better cost) on the bottom. The numerical value of the cost is indicated above each panel.}
    \label{fig:experimental_preferred}
    \vspace{-12pt}
\end{figure*}

\begin{figure}[b]
    \centering
    \includegraphics[width=\columnwidth]{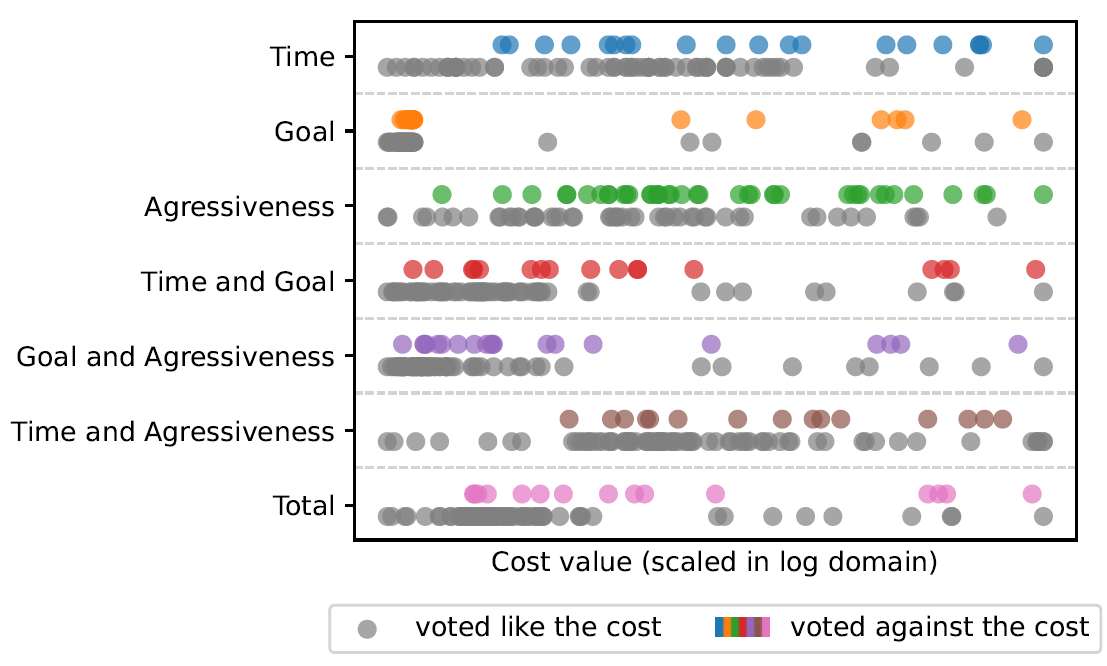}
    \vspace{-15pt}
    \caption{Values of the different costs of the selected algorithms. Coloured marks indicate experiments where human preference differed from the cost function. }
    \vspace{-10pt}
    \label{fig:experimental_cost_preference}
\end{figure}

The human aspect of PBO is often overlooked in the literature. In most studies, comparisons are performed using predefined cost functions rather than human input, to enable consistent and reproducible evaluation across algorithms. However, as discussed in Section~\ref{sec:introduction}, a human expert possesses multimodal sensory information and an internal mental model shaped by experience, factors that an algorithm cannot fully replicate. What may appear as inconsistency in human preference judgments could instead reflect the complexity of the latent objective function in the expert’s mind, which may integrate multiple, often non-quantifiable, criteria.

\newpage

To investigate these effects, we evaluate the refined cost for each experiment conducted during the expert-guided trials. This allows us to compare the actual selections of the expert with the corresponding cost values they would have produced under the refined cost. As shown in Fig.~\ref{fig:experimental_preferred}, several instances reveal the expert favouring the upper experiment despite its higher adapted cost, suggesting that defining a cost function fully aligned with human preferences is non-trivial. We therefore analyse in more detail the cases where the expert’s choices deviate from the cost-based predictions to identify potential systematic patterns.

Fig.~\ref{fig:experimental_cost_preference} illustrates the refined cost values for each experiment, grouped row-wise by cost component. Coloured markers denote instances where the expert’s choice differed from the cost-based prediction, while grey markers indicate agreement. The individual cost terms \textit{Time}, \textit{Goal}, and \textit{Aggressiveness} correspond to $J_1$, $J_2$, and $J_3$ in Eq.~\ref{eq:cost_combination}, and the remaining rows show combinations of these components (subsets of the overall cost function). We observe that the leftmost 10\% of data points—representing experiments with the lowest (most favourable) costs—are predominantly grey, indicating that the expert rarely disagreed with the cost function in these cases. Beyond this region, the number of agreements and disagreements for the goal-related cost becomes roughly balanced, implying that once the goal cost exceeds a certain threshold, the expert begins to weigh other factors more heavily.

\begin{figure}[h]
    \centering
    \includegraphics[width=\columnwidth]{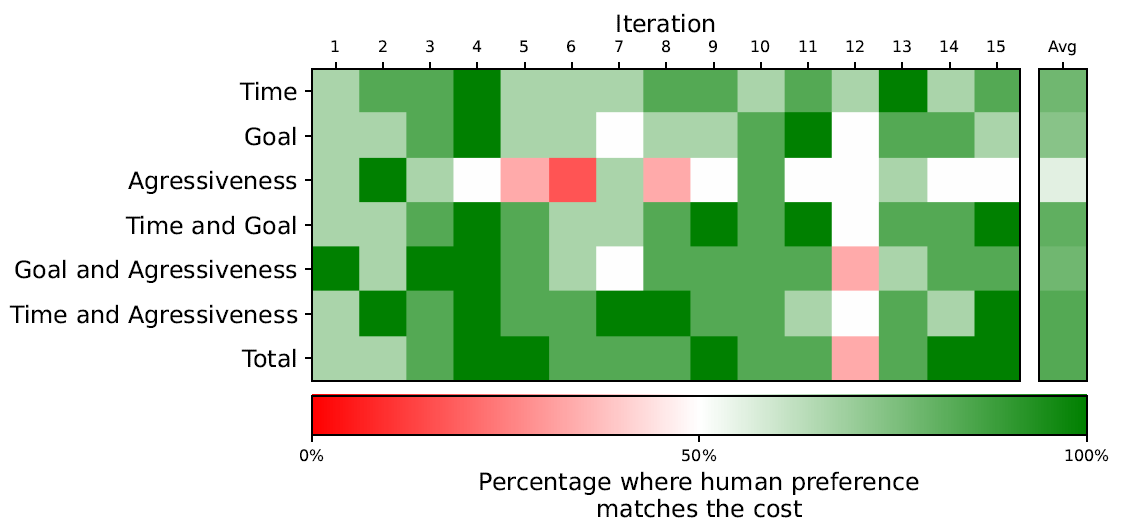}
    \vspace{-18pt}
    \caption{Percentage per iteration that the human-selected experiments agreed with the designed cost function, indicating that human preference is not well modelled by a simple cost.}
    \vspace{-10pt}
    \label{fig:experimental_time_preference}
\end{figure}

Another possible pattern may emerge when analysing the expert’s selections over successive iterations. It seems plausible that the expert emphasizes different cost terms at different stages; for instance, prioritizing one criterion during early iterations and shifting focus to another later on. As shown in Fig.~\ref{fig:experimental_time_preference}, aggressiveness was generally weighted less heavily than the other cost terms, while the total cost was followed relatively consistently throughout the experiments. However, no clear or repeatable pattern could be identified that would enable a practical modelling of this behaviour. Overall, these observations underscore the difficulty for an expert to design a cost function that accurately reflects their own decision-making process.

\subsection{Data-driven cost function from expert preferences}
The fact that no clear pattern could be identified in the expert’s selections motivates us to adopt an alternative approach to defining an expert-guided cost function. An inherent by-product of most PBO algorithms is an internal Gaussian Process (GP) model that maps the control parameters to an abstract cost representation. Although this cost has no direct physical interpretation, it provides a smooth, data-driven estimate of the expert’s underlying preferences: parameter settings predicted to have the lowest GP cost correspond to those most preferred by the expert. This learned cost model can subsequently be employed with standard cost-based tuning methods to automate parameter selection without requiring further human input. While any optimization algorithm could in principle be used, we employ Bayesian Optimization (BO) for consistency with the earlier PBO experiments and to enable a fair, qualitative comparison for the expert.

\begin{figure}[h]
    \centering
    \includegraphics[width=\columnwidth]{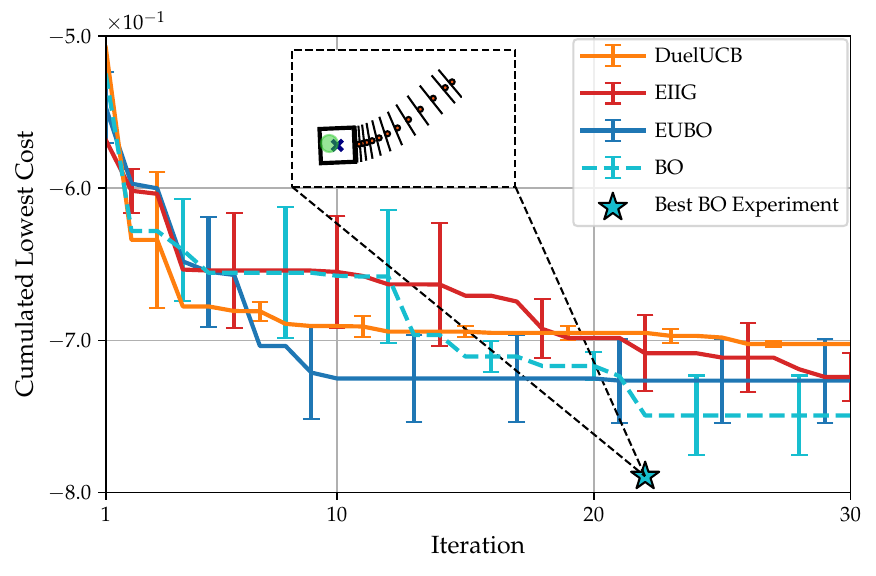}
    \vspace{-18pt}
    \caption{Average and standard error of the cumulative minimum cost obtained from experiments on the robotic setup. The y-axis represents the data-driven Gaussian Process (GP) cost. The experiments of Bayesian Optimization (\textit{BO}) use the GP cost, while the other lines show the experiments performed by expert preference selection from Section \ref{sec:PBO_setup}, but remapped to also show the GP cost for those experiments. Note, BO has access to a cost function, while the others only used expert preferences.}
    \vspace{-6pt}
    \label{fig:bo_benchmark}
\end{figure}

To construct this expert-guided cost function, all pairwise preference data collected from the expert (see Fig.~\ref{fig:experimental_benchmark}) were combined to train a Pairwise Gaussian Process model, similar to the internal surrogate used within PBO. This model provides a continuous estimate of the latent cost surface that best explains the observed preferences. Once trained, the GP is kept fixed and serves as a static cost function that maps any controller parameter vector directly to a scalar cost value; no additional experiments or updates are performed during optimization. To verify this interpretation, we applied Bayesian Optimization to this learned cost surface. This step effectively tests whether the extracted cost model captures the expert’s decision process: BO explores the parameter space automatically using the GP cost as a guide. The resulting performance, shown in Fig.~\ref{fig:bo_benchmark}, demonstrates that optimization over this data-driven cost yields trajectories that align closely with the expert’s preferences and even surpass the original preference-based tuning when evaluated on the same metric.

\begin{figure}[h]
    \centering
    \includegraphics[width=\columnwidth]{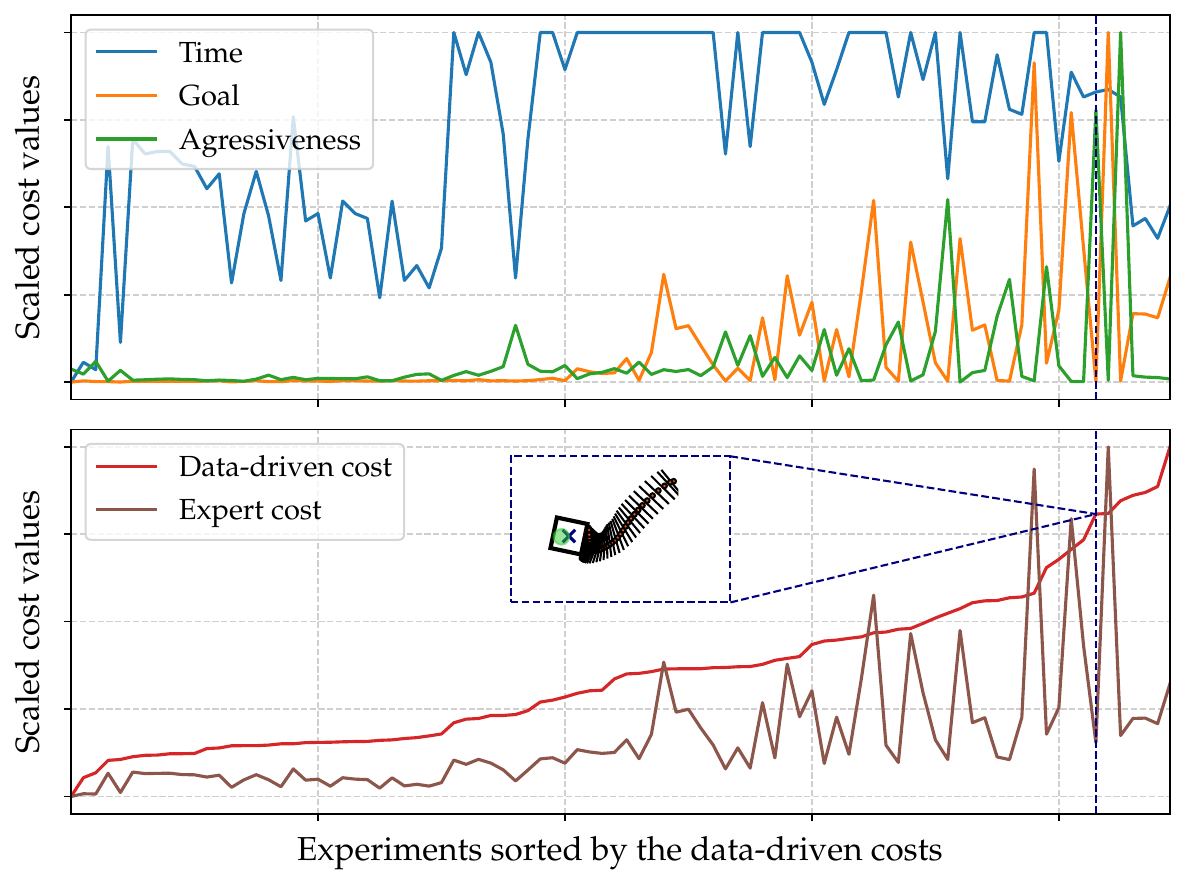}
    \vspace{-15pt}
    \caption{The experiments of Bayesian Optimization sorted according to the data-driven cost. It shows the relative importance it gives to the three expert defined cost terms (Time, Goal and Aggressiveness) and their total (expert-driven cost).}
    \vspace{-8pt}
    \label{fig:cost_comparisson_experimental_bo}
\end{figure}

Each BO run was subsequently validated on the real setup, confirming that the expert’s qualitative assessments aligned with the predictions of the data-driven cost function. This agreement supports the interpretation that the GP-based model provides a faithful and quantitative representation of the expert’s preferences, which in turn enables a direct comparison between the expert-defined and data-driven cost formulations. As shown in Fig.~\ref{fig:cost_comparisson_experimental_bo}, all trials with lower data-driven cost successfully reached the goal. Lower cost values also clearly correspond to less aggressive motions. However, for parameter ranges where the data-driven cost results in a higher value, the contribution of the individual cost components becomes less distinct, particularly the influence of time, which appears less straightforward than initially expected. One trajectory, for instance, exhibits a relatively low expert-defined cost but a high data-driven cost. This push is executed rather slowly and requires a large corrective motion at the end, which is a behaviour that is not yet penalized in the expert-defined cost function.

\newpage

Overall, these findings indicate that expert preference selections can be used to infer a cost function suitable for automated tuning, potentially reducing the need for repeated manual intervention.

\section{Conclusions and Future Research}
In this work, we investigated the use of Preferential Bayesian Optimization (PBO) to support the tuning of a manipulator tasked with pushing a block to a target position. We benchmarked multiple state-of-the-art PBO algorithms and demonstrated that they are capable of effectively guiding the system toward favourable settings both in simulation and in real-world experiments.

We further examined whether human preferences could be adequately captured by an expert-designed cost function. Our study revealed that, in practice, experts are unable to define a cost function that fully aligns with their own decision-making process. To address this limitation, we showed that an implicit, data-driven cost function—constructed as a by-product of the PBO algorithm—can better reflect human judgments. By doing so, the resulting objective best represents the expert’s preferences, which could be used for standard cost-based Bayesian Optimization.

Despite these positive results, several practical limitations remain for existing PBO algorithms, pointing to directions for future research. For example, performance is often measured as a function of the number of comparisons made so far. We argue that the total number of experiments required is a more meaningful metric. This observation motivates further investigation into richer preference structures, such as full orderings instead of simple pairwise duels, or strategies that allow comparisons between newly proposed experiments and multiple previous trials to enhance exploration.

Another shortcoming we observed is that the exploration-exploitation trade-off is often neglected. Allowing the operator to dynamically adjust this trade-off based on observed results could improve the efficiency and effectiveness of the optimization.


\section*{Acknowledgment}
This work was supported by the Flanders Make project REXPEK and the Research Foundation Flanders (FWO) project CTRLxAI (SBO grant no. S007723N).

\bibliographystyle{unsrt}
\bibliography{Refs}

\begin{thebibliography}{10}

\bibitem{ericsson1993role}
K~Anders Ericsson, Ralf~T Krampe, and Clemens Tesch-R{\"o}mer.
\newblock The role of deliberate practice in the acquisition of expert performance.
\newblock {\em Psychological review}, 100(3):363, 1993.

\bibitem{chase1973chess}
William~G. Chase and Herbert~A. Simon.
\newblock The mind's eye in chess.
\newblock In William~G. Chase, editor, {\em Visual Information Processing}, pages 215--281. Academic Press, 1973.

\bibitem{lobo2025healthcare}
Afonso Lobo, Daniel S{\'a}, Jo{\~a}o Cunha, Ricardo Duarte, J{\'u}lio Duarte, and Manuel~Filipe Santos.
\newblock Healthcare cognitive decision support system--the state-of-the-art.
\newblock {\em Procedia Computer Science}, 257:1104--1109, 2025.

\bibitem{gerlich2025ai}
Michael Gerlich.
\newblock {AI} tools in society: Impacts on cognitive offloading and the future of critical thinking.
\newblock {\em Societies}, 15(1):6, 2025.

\bibitem{hao2021construction}
Xuejie Hao, Zheng Ji, Xiuhong Li, Lizeyan Yin, Lu~Liu, Meiying Sun, Qiang Liu, and Rongjin Yang.
\newblock Construction and application of a knowledge graph.
\newblock {\em Remote Sensing}, 13(13):2511, 2021.

\bibitem{nonaka2009knowledge}
Ikujiro Nonaka.
\newblock The knowledge-creating company.
\newblock In {\em The economic impact of knowledge}, pages 175--187. Routledge, 2009.

\bibitem{mockus2005bayesian}
Jonas Mockus.
\newblock The {Bayesian} approach to global optimization.
\newblock In {\em System Modeling and Optimization: Proceedings of the 10th IFIP Conference New York City, USA, August 31--September 4, 1981}, pages 473--481. Springer, 2005.

\bibitem{jones1998efficient}
Donald~R Jones, Matthias Schonlau, and William~J Welch.
\newblock Efficient global optimization of expensive black-box functions.
\newblock {\em Journal of Global optimization}, 13(4):455--492, 1998.

\bibitem{shahriari2015taking}
Bobak Shahriari, Kevin Swersky, Ziyu Wang, Ryan~P Adams, and Nando De~Freitas.
\newblock Taking the human out of the loop: A review of bayesian optimization.
\newblock {\em Proceedings of the IEEE}, 104(1):148--175, 2015.

\bibitem{greenhill2020bayesian}
Stewart Greenhill, Santu Rana, Sunil Gupta, Pratibha Vellanki, and Svetha Venkatesh.
\newblock {Bayesian} optimization for adaptive experimental design: A review.
\newblock {\em IEEE access}, 8:13937--13948, 2020.

\bibitem{coutinho2023bayesian}
Jo{\~a}o~PL Coutinho, Lino~O Santos, and Marco~S Reis.
\newblock {Bayesian} optimization for automatic tuning of digital multi-loop {PID} controllers.
\newblock {\em Computers \& Chemical Engineering}, 173:108211, 2023.

\bibitem{neumann2019data}
Matthias Neumann-Brosig, Alonso Marco, Dieter Schwarzmann, and Sebastian Trimpe.
\newblock Data-efficient autotuning with {Bayesian} optimization: An industrial control study.
\newblock {\em IEEE Transactions on Control Systems Technology}, 28(3):730--740, 2019.

\bibitem{konig2023safe}
Christopher K{\"o}nig, Miks Ozols, Anastasia Makarova, Efe~C Balta, Andreas Krause, and Alisa Rupenyan.
\newblock Safe risk-averse {Bayesian} optimization for controller tuning.
\newblock {\em IEEE Robotics and Automation Letters}, 8(12):8208--8215, 2023.

\bibitem{konig2025adaptive}
Christopher K{\"o}nig, Raamadaas Krishnadas, Efe~C Balta, and Alisa Rupenyan.
\newblock Adaptive {Bayesian} optimization for high-precision motion systems.
\newblock {\em IEEE Transactions on Automation Science and Engineering}, 2025.

\bibitem{letham2019bayesian}
Benjamin Letham and Eytan Bakshy.
\newblock {Bayesian} optimization for policy search via online-offline experimentation.
\newblock {\em Journal of Machine Learning Research}, 20(145):1--30, 2019.

\bibitem{eugene2023learning}
Elvis~A Eugene, Kyla~D Jones, Xian Gao, Jialu Wang, and Alexander~W Dowling.
\newblock Learning and optimization under epistemic uncertainty with {Bayesian} hybrid models.
\newblock {\em Computers \& Chemical Engineering}, 179:108430, 2023.

\bibitem{sabbatella2024bayesian}
Antonio Sabbatella, Andrea Ponti, Antonio Candelieri, and Francesco Archetti.
\newblock {Bayesian} optimization using simulation-based multiple information sources over combinatorial structures.
\newblock {\em Machine Learning and Knowledge Extraction}, 6(4):2232--2247, 2024.

\bibitem{karkaria2024towards}
Vispi Karkaria, Anthony Goeckner, Rujing Zha, Jie Chen, Jianjing Zhang, Qi~Zhu, Jian Cao, Robert~X Gao, and Wei Chen.
\newblock Towards a digital twin framework in additive manufacturing: Machine learning and {Bayesian} optimization for time series process optimization.
\newblock {\em Journal of Manufacturing Systems}, 75:322--332, 2024.

\bibitem{gafurov2025ai}
Anton~Nailevich Gafurov, Sooyoung Lee, Uzair Ali, Muhammad Irfan, Inyoung Kim, and Taik-Min Lee.
\newblock {AI}-driven digital twin for autonomous web tension control in roll-to-roll manufacturing system.
\newblock {\em Scientific Reports}, 15(1):1--17, 2025.

\bibitem{nobar2024guided}
Mahdi Nobar, J{\"u}rg Keller, Alisa Rupenyan, Mohammad Khosravi, and John Lygeros.
\newblock Guided {Bayesian} optimization: Data-efficient controller tuning with digital twin.
\newblock {\em IEEE Transactions on Automation Science and Engineering}, 2024.

\bibitem{EP_TS_MUC}
Javier Gonz{\'a}lez, Zhenwen Dai, Andreas Damianou, and Neil~D Lawrence.
\newblock Preferential {Bayesian} optimization.
\newblock In {\em International Conference on Machine Learning}, pages 1282--1291. PMLR, 2017.

\bibitem{chu2005preference}
Wei Chu and Zoubin Ghahramani.
\newblock Preference learning with {Gaussian} processes.
\newblock In {\em Proceedings of the 22nd international conference on Machine learning}, pages 137--144, 2005.

\bibitem{maccarini2022preference}
Marco Maccarini, Filippo Pura, Dario Piga, Loris Roveda, Lorenzo Mantovani, and Francesco Braghin.
\newblock Preference-based optimization of a human-robot collaborative controller.
\newblock {\em IFAC-PapersOnLine}, 55(38):7--12, 2022.

\bibitem{tucker2022polar}
Maegan Tucker, Kejun Li, Yisong Yue, and Aaron~D Ames.
\newblock Polar: Preference optimization and learning algorithms for robotics.
\newblock {\em arXiv preprint arXiv:2208.04404}, 2022.

\bibitem{LA_EI}
Eric Brochu, Vlad~M Cora, and Nando De~Freitas.
\newblock A tutorial on {Bayesian} optimization of expensive cost functions, with application to active user modeling and hierarchical reinforcement learning.
\newblock {\em arXiv preprint arXiv:1012.2599}, 2010.

\bibitem{EP_MUC}
Tristan Fauvel and Matthew Chalk.
\newblock Efficient exploration in binary and preferential {Bayesian} optimization.
\newblock {\em arXiv preprint arXiv:2110.09361}, 2021.

\bibitem{DuelTSUCB}
Alessio Benavoli, Dario Azzimonti, and Dario Piga.
\newblock Preferential bayesian optimisation with skew {Gaussian} processes.
\newblock In {\em Proceedings of the Genetic and Evolutionary Computation Conference Companion}, pages 1842--1850, 2021.

\bibitem{EUBO}
Zhiyuan~Jerry Lin, Raul Astudillo, Peter Frazier, and Eytan Bakshy.
\newblock Preference exploration for efficient {Bayesian} optimization with multiple outcomes.
\newblock In {\em International Conference on Artificial Intelligence and Statistics}, pages 4235--4258. PMLR, 2022.

\bibitem{HB}
Shion Takeno, Masahiro Nomura, and Masayuki Karasuyama.
\newblock Towards practical preferential {Bayesian} optimization with skew {Gaussian} processes.
\newblock In {\em International Conference on Machine Learning}, pages 33516--33533. PMLR, 2023.

\bibitem{dewitte2025}
Sander~De Witte, Tom Lefebvre, Thomas Neve, Andras Retzler, and Guillaume Crevecoeur.
\newblock Differential flatness of quasi-static slider-pusher models with applications in control, 2025.

\end{thebibliography}

\end{document}